\documentclass[twocolumn]{article}
\usepackage[margin=0.6in,columnsep=0.15in]{geometry}
\usepackage[utf8]{inputenc}

\usepackage[usenames,dvipsnames]{xcolor}
\usepackage[colorlinks,citecolor=Gray,urlcolor=Gray,linkcolor=Blue]{hyperref}

\usepackage{graphicx}
\usepackage{mathtools}
\usepackage{booktabs}
\usepackage{tabularx}
\usepackage{tikz}
\usepackage{siunitx}
\usepackage{dcolumn}

\usepackage[lcgreekalpha,upint]{stix}

\usepackage[tiny]{titlesec}
\usepackage[inline]{enumitem}

\usepackage{algorithm} 
\usepackage{algpseudocode} 

\usepackage[normal]{caption}
\DeclareCaptionLabelSeparator{pipe}{ $|$ }
\usepackage{authblk}

\usepackage[square,sort&compress,comma,numbers]{natbib}

\usepackage{doi}

\usepackage{subfigure}

\usepackage{hyperref}

\usepackage[switch]{lineno}
\title{\Large \textit{Ab initio} calculation of real solids via neural network ansatz }

\author[1*]{Xiang Li}
\author[1]{Zhe Li}
\author[2]{Ji Chen}
\affil[1]{ByteDance Inc, Zhonghang Plaza, No. 43,  North 3rd Ring West Road, Haidian District, Beijing.}
\affil[2]{School of Physics, Interdisciplinary Institute of Light-Element Quantum Materials, Frontiers
Science Center for Nano-Optoelectronics, Peking University, Beijing 100871, People’s Republic of China}

\date{}

\makeatletter
\def\blfootnote{\xdef\@thefnmark{}\@footnotetext}
\makeatother

\begin{document}

\twocolumn[{%
  \maketitle
  \vspace{-3em}
  \begin{center}
  \begin{minipage}{0.85\linewidth}
    \small
    \paragraph{Abstract}

    Neural networks have been applied to tackle many-body electron correlations for small molecules and physical models in recent years.
    Here we propose a new architecture that extends molecular neural networks with the inclusion of periodic boundary conditions to enable \textit{ab initio} calculation of real solids.
    %
    The accuracy of our approach is demonstrated in four different types of systems, namely the one-dimensional periodic hydrogen chain, the two-dimensional graphene, the three-dimensional lithium hydride crystal, and the homogeneous electron gas, 
    where the obtained results, e.g. total energies, dissociation curves, and cohesive energies, outperform many traditional \textit{ab initio} methods and reach the level of the most accurate approaches.
    Moreover, electron densities of typical systems are also calculated to provide physical intuition of various solids.
    Our method of extending a molecular neural network to periodic systems can be easily integrated into other neural network structures, highlighting  
    %
    a promising future of \textit{ab initio} solution of more complex solid systems using neural network ansatz, and more generally endorsing the application of machine learning in materials simulation and condensed matter physics.
    
  \end{minipage}
  \end{center}
  \vspace{1em}
}]

\blfootnote{$^*$Email: lixiang.62770689@bytedance.com}%


\let\oldAA\AA
\renewcommand{\AA}{\text{\normalfont\oldAA}}
\newcommand{\red}[1]{\textcolor{red}{#1}}
\newcommand{\bx}{\mathbf{x}}
\newcommand{\br}{\mathbf{r}}
\newcommand{\bk}{\mathbf{k}}
\newcommand{\bR}{\mathbf{R}}
\newcommand{\bL}{\mathbf{L}}
\newcommand{\bg}{\mathbf{g}}
\newcommand{\bG}{\mathbf{G}}
\newcommand{\bom}{\mathbf{\Omega}}

\hyphenation{Schrö-din-ger}




\section{Introduction}

Solving the many-body electronic structure of real solids from \textit{ab initio} is one of the grand challenges in condensed matter physics and materials science \citep{KohnNobel}. 
More accurate \textit{ab initio} solutions can push the limit of our understanding in many fundamental and mysterious emergent phenomena, such as superconductivity, light-matter interaction, heterogeneous catalysis, to name just a few \citep{martin_2004}. 
%
The current workhorse method 
is density functional theory (DFT), whose accuracy 
depends quite sensitively on the choice of the so-called exchange-correlation functional and unfortunately there lacks a systematic routine towards the exact \citep{DFT-JONES,kirkpatrick_pushing_2021}.
Other commonly used \textit{ab initio} quantum chemistry methods, such as the coupled-cluster and configuration interaction theories \cite{simons2020}, can provide more accurate solutions for molecules but face severe difficulty when applied to solid systems.
Recently, several breakthroughs have been made in applying these quantum chemistry methods on solids \citep{Booth2013TowardsAE,mihm_shortcut_2021}, driving the study of solid systems towards a new frontier.

Meanwhile, in the last few years, new attempts to tackle the correlated wavefunction problem in molecules or model Hamiltonians using neural network based approaches have been reported by different groups \citep{han_solving_2019,RBM_ising,FermiNet,spencer2020better,PauliNet,ferminetecp, ferminet_dmc}. 
The key idea is to use the neural network as the wavefunction ansatz in quantum Monte Carlo (QMC) simulations.
%
The stochastic nature of QMC enables a considerably economical 
\citep{guther_neci_2020,FoulkesReview,ShiAFQMC,Booth2013TowardsAE} time scaling and efficient parallelization.
The universal approximation theorem behind neural network based ansatz significantly improves the accuracy of the traditional QMC method and the strategy has been proved successful in studying small molecules \cite{FermiNet,spencer2020better,PauliNet}.
However, how to apply such neural network ansatz for real solids, i.e. how to apply periodic boundary conditions (PBC) in the neural network, and whether it can describe the long-range electron correlations in extended systems remain as open questions.

Here we propose a powerful periodic neural network ansatz, based on which we develop a highly efficient QMC method for \textit{ab initio} calculation of real solid and general periodic systems with unprecedented accuracy.
We apply our method to periodic hydrogen chains, graphene, lithium hydride (LiH) crystal, and homogeneous electron gas.
These systems cover a wide range of interests, including materials dimension from one to three, electronic structure from metallic to insulating, and bonding type from covalent to ionic. 
The calculated dissociation curve, cohesive energy and correlation energy, can be compared satisfactorily with available experimental values and other state-of-the-art computational approaches. 
Electron densities of typical systems are further calculated to test our neural network and explore the underlying physics.  
All the results demonstrates that our method can achieve accurate electronic structure calculations of real solid/periodic systems. 

\section{Results}  

\subsection{Neural network for solid system}
\begin{figure*}[t]
    \centering
    \includegraphics[width=2.0\columnwidth]{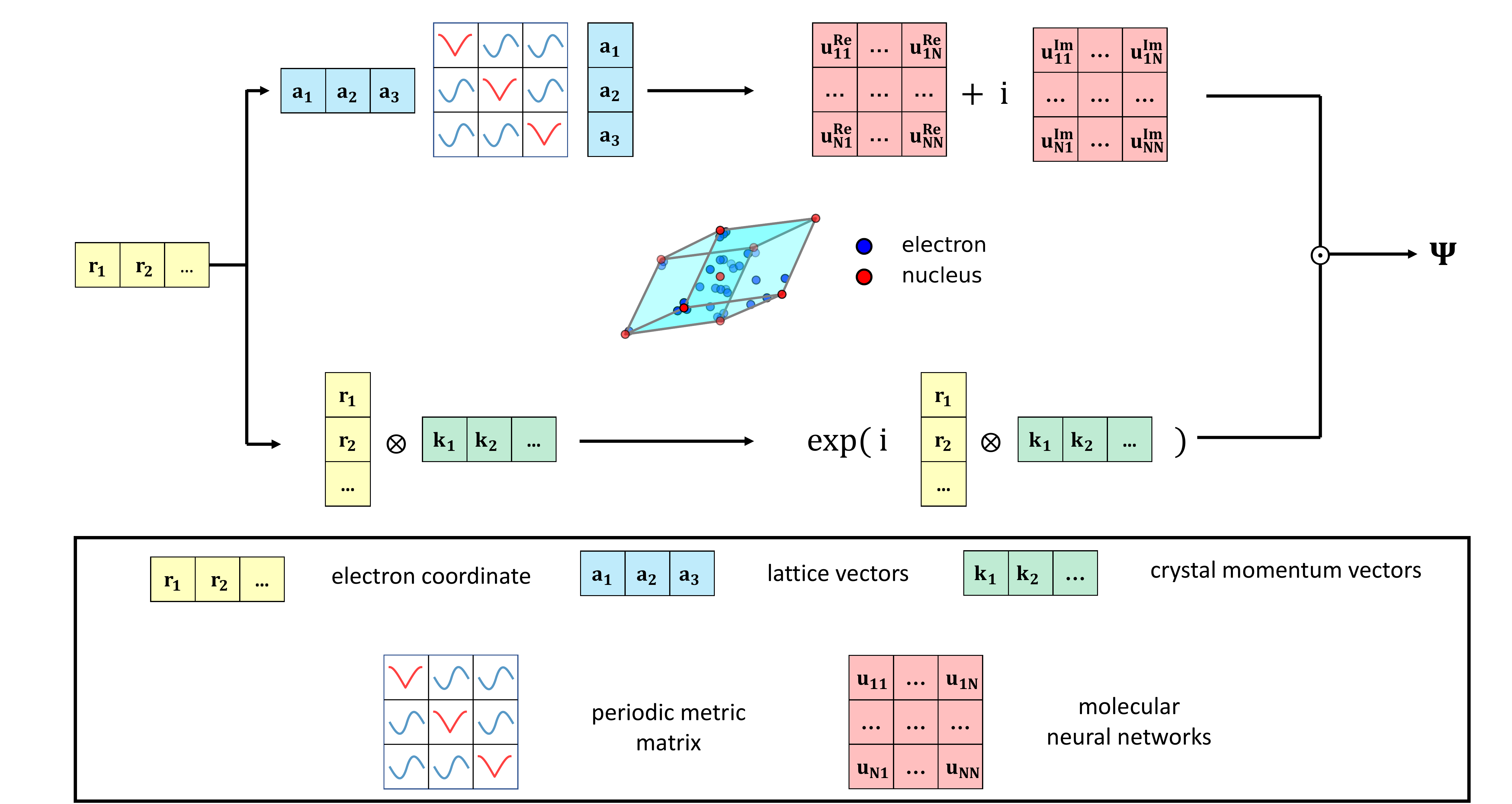}
    \caption{\textbf{Sketch of neural network architecture.} The electron coordinates ${\br_i}$ are passed to two channels. In the first one, they build the periodic distance features $d(\br)$ using the periodic metric matrix $\mathbf{M}$ and the lattice vectors ${\mathbf{a}}$, and then $d(\br)$ features are fed into two molecular neural networks, that represent separately the real and the imaginary part of wavefunction. In the second channel, ${\br_i}$ constructs the plane-wave phase factors on a selected set of crystal momentum vectors. The total wavefunctions of solids are constructed by the two channels following the expression of Eq.~\eqref{eq:ansatz}.}
    \label{fig:net}
\end{figure*}

Periodicity and anti-symmetry are two fundamental properties of the wavefunction of a solid system. 
The anti-symmetry can be ensured by the Slater determinant, which has been commonly used as the basic block in molecular neural networks. 
We also express the wavefunction by two Slater determinants of one spin-up channel and one spin-down channel,

\begin{equation}
  \Psi(\br) = {\rm Det}_\uparrow[e^{i\bk\cdot \br_\uparrow}u^\uparrow_{\rm mol}(d)]{\rm Det}_\downarrow[e^{i\bk\cdot \br_\downarrow}u_{\rm mol}^\downarrow(d)]\ .
  \label{eq:ansatz}
\end{equation}
In this regard, our ansatz resembles the structure of FermiNet \citep{FermiNet,spencer2020better}, whereas other neural network wavefunction ansatz may include extra terms in addition to the Slater determinants \citep{PauliNet}.  
Each determinant is then constructed from a set of periodic orbitals, which inherits the physics captured by the Bloch function form by a product of phase factor $e^{i\bk\cdot \br}$ and collective molecular orbital $u_{\rm mol}$.

Fig.~\ref{fig:net} displays more details on the structure of our neural network.
Building an efficient and accurate periodic ansatz is the key step in developing \textit{ab initio} methods for solid. 
Here we have followed the recently proposed scheme of Whitehead et al. to construct a set of periodic distance features  $d(\br)$ \citep{PeriodicDis} using lattice vectors in real and reciprocal space $(\mathbf{a}_i,\mathbf{b}_i)$,
\begin{equation}
\begin{gathered}
    d(\br)=\frac{\sqrt{\mathbf{A}\mathbf{M}\mathbf{A}^T}}{2\pi}~,~
    \mathbf{A}=(\mathbf{a}_1,\mathbf{a}_2,\mathbf{a}_3)\ ,\\
    \mathbf{M}_{ij}=f^2(\omega_i)\delta_{ij}+g(\omega_i)g(\omega_j)(1-\delta_{ij})
    ~,~\omega_i=\br\cdot\mathbf{b}_i\ .
\end{gathered}
\label{eq:periodic_metric}
\end{equation}
The periodic metric matrix $\mathbf{M}$ is constructed by periodic functions $f,g$, which retains ordinary distances at the origin and regulates them to periodic ones at far distances, ensuring asymptotic cusp form, continuity, and periodicity requirement at the same time.

The constructed periodic distance features $d(\br)$ can then be fed into molecular neural networks to form collective orbitals $u_{\rm mol}$. Specifically, in this work we represent the molecular networks with FermiNet \citep{FermiNet}, which incorporates the electron-electron interactions. 
The inclusion of all-electron features promotes the traditional single-particle orbitals to the collective ones, 
and hence the description of wavefunction and correlation effects can be improved while fewer Slater determinants are required.
In addition, the wavefunction of solid systems is necessarily complex-valued, 
and we introduce two sets of molecular orbitals 
to represent the real and imaginary parts of the solid wavefunction, respectively.
The plane-wave phase factors $e^{{\rm i}\bk\cdot\br}$ in Fig.~\ref{fig:net} are used to construct the Bloch function like orbitals, and
the corresponding $\bk$ points are selected to minimize the Hartree-Fock (HF) energy.


Based on the variational principle, our neural network is trained using the variational Monte Carlo (VMC) approach. To efficiently optimize the network, a modified Kronecker factored curvature estimator (KFAC) optimizer \citep{kfac} is adopted,
which significantly outperforms traditional energy minimization algorithms. 
Calculations are also ensured by efficient and massive parallelization on multiple nodes of high-performance GPUs. 
More details on the theories, methods, and computations are included in the Methods section and the supplementary information.

\begin{figure*}[t]
    \centering
    \includegraphics[width=2.0\columnwidth]{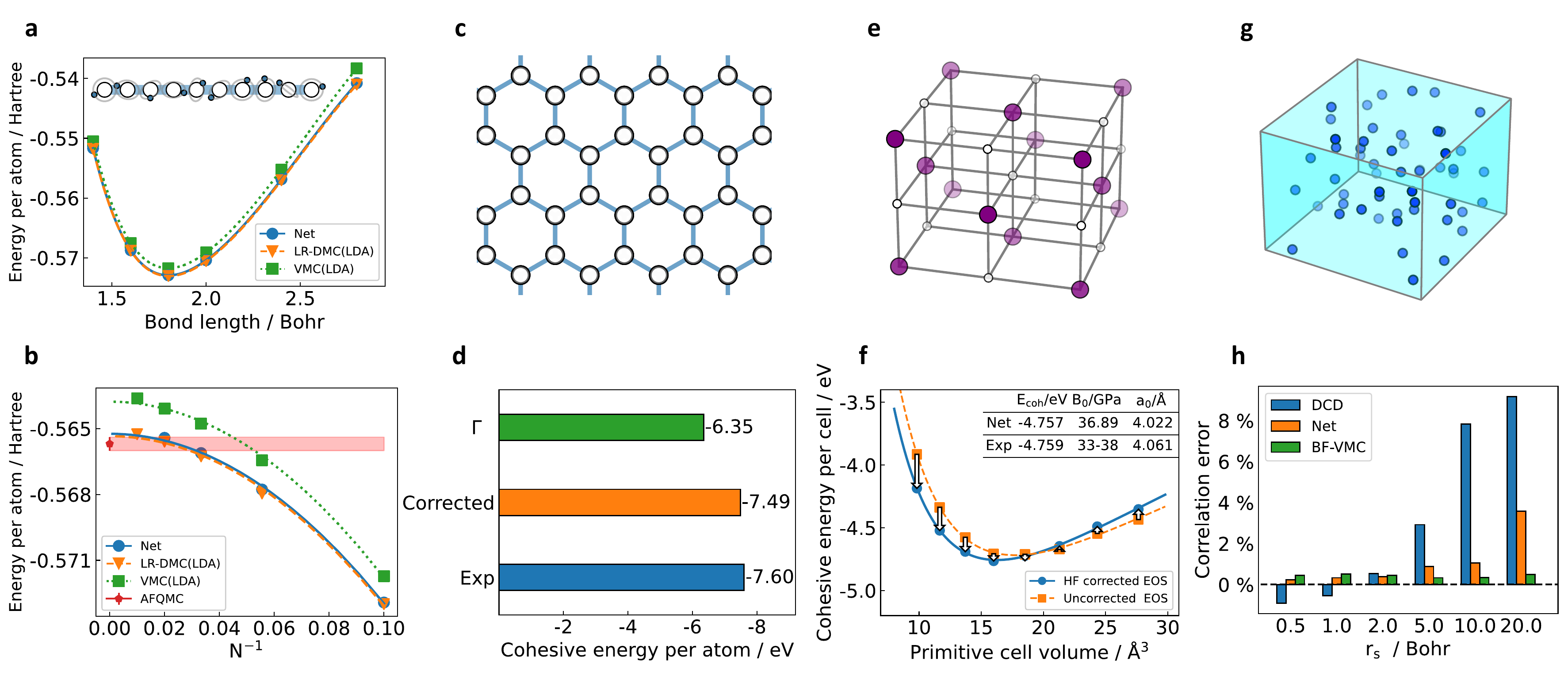}
	\caption{\textbf{Calculated results of neural network}. Our results are all labeled as Net.
    ${\rm \mathbf{a}}$, ${\rm H}_{10}$ dissociation curve is plotted.
    ${\rm \mathbf{b}}$, energy of different hydrogen chain sizes N, the bond length of hydrogen chain is fixed at 1.8 Bohr. LR-DMC and VMC both use TZ-LDA basis and AFQMC is pushed to complete basis limit \citep{hydrogen_chain}.
    ${\rm \mathbf{c}}$, structure of graphene.
    ${\rm \mathbf{d}}$, the cohesive energy per atom of $\Gamma$ point and finite-size error corrected result is plotted. Experiment cohesive energy is from Ref.~\cite{Graphene_exp}. Graphene is calculated at its equilibrium length 1.421 \AA.
    ${\rm \mathbf{e}}$, structure of rock-salt lithium hydride crystal.
    ${\rm \mathbf{f}}$, equation of state of LiH crystal is plotted, fitted Birch-Murnaghan parameters and experimental data are also given. HF corrections are calculated using ccpvdz basis, and $E_\infty^{\rm HF}$ is approximated by $E_{\rm N=8}^{\rm HF}$. The arrows denote the corresponding HF corrections. 
    ${\rm \mathbf{g}}$, plot of homogeneous electron gas system.
    ${\rm \mathbf{h}}$, correlation error of 54 electrons HEG systems at different $r_s$. DCD and BF-VMC results are displayed for comparison, and BF-DMC data is used as reference \citep{heg_method_3,heg_method_5}.}
    \label{fig:unify}
\end{figure*}

\subsection{Ground-state energy}
\subsubsection{Hydrogen chain} 
Hydrogen chain is one of the simplest models in condensed matter research. Despite its simplicity, hydrogen chain is a challenging and interesting system, serving as a benchmark system for electronic structure methods and featuring intriguing correlated phenomena \citep{hydrogen_chain}.
The calculated energy of the periodic $\text{H}_{10}$ chain as a function of the bond length is shown in Fig.~\ref{fig:unify}a.
The results from lattice-regularized diffusion Monte Carlo (LR-DMC) and
traditional VMC 
are also plotted for comparison \citep{hydrogen_chain}. 
We can see that our results nearly coincide with the LR-DMC results and significantly outperform traditional VMC (see Supplementary Table 3). 
In Fig.~\ref{fig:unify}b, the energy of hydrogen chains of different atom numbers are calculated for extrapolation to the
thermodynamic limit (TDL). 
The shaded bar in Fig.~\ref{fig:unify}b illustrates the extrapolated energy of the periodic hydrogen chain at TDL from auxiliary field quantum Monte Carlo (AFQMC), which is considered as the current state-of-the-art along with LR-DMC.
Our TDL result is comparable with both AFQMC and LR-DMC (see Supplementary Table 4).


\subsubsection{Graphene} 
Graphene is arguably the most famous two-dimensional system (Fig.~\ref{fig:unify}c)
receiving broad attention in the past two decades for its mechanical, electronic, and chemical applications \citep{geim_nobel}.
Here we carry out simulations
to estimate its cohesive energy, which measures the strength of C-C chemical bonding and long-range dispersion interactions.
The calculations are performed on a $2\times2$ supercell of graphene using twist average boundary condition (TABC) \citep{twist_average} in conjunction with structure factor $S(\bk)$ correction \citep{sf_correction} (see Supplementary Fig. 3) to 
reduce the finite-size error. 
The calculated results are plotted in Fig.~\ref{fig:unify}d along with the experimental value \cite{Graphene_exp}, and it shows that our neural network can deal with graphene very well, producing a cohesive energy of graphene within 0.1 eV/atom to the experimental reference (see Supplementary Table 6). 
We also plotted the results with PBC, 
namely the $\Gamma$ point only result, which deviates from the experiment data by 1.25 eV/atom.

\subsubsection{Lithium hydride crystal} 
For three-dimensional system, we consider the LiH crystal with a rock-salt structure (Fig.~\ref{fig:unify}e),
another benchmark system for accurate \textit{ab initio} methods \citep{Booth2013TowardsAE, scf_lih, qmc_lih}.
Despite consisting of only simple elements, LiH represents typical ionic and covalent bondings upon changing the lattice constants.
Using our neural network, we first simulate the equation of state of LiH on a $2\times2\times2$ supercell, as shown in Fig.~\ref{fig:unify}f.
%
In addition, we employ a standard finite-size correction based on Hartree-Fock calculations of a large supercell (see Supplementary Fig. 5).
In Fig.~\ref{fig:unify}f we also show the Birch-Murnaghan fitting to the equation of state, based on which we can obtain thermodynamic quantities such as the cohesive energy, the bulk modulus, and the equilibrium lattice constant of LiH.
As shown in the inset, our results on the thermodynamic quantities agree very well with experimental data \citep{scf_lih} (see Supplementary Table 8, 9).
%
For further validation, we have also computed directly the $3\times3\times3$ supercell of LiH at its equilibrium length $4.061 \AA$, which contains 108 electrons. 
To the best of our knowledge, this is the largest electronic system computed using a high-quality neural network ansatz. The $3\times3\times3$ supercell calculation predicts the total energy per unit cell of LiH is $-8.160$ Hartree and the cohesive energy per unit cell is $-4.770~{\rm eV}$ after the finite-size correction (see Supplementary Table 10), which is also very close to the experimental value $-4.759~{\rm eV}$ \citep{scf_lih}.



\subsubsection{Homogeneous electron gas}
In addition to the real solids, our computational framework can also apply straightforwardly to model systems such as homogeneous electron gas (HEG).
HEG has been studied for a long time to understand the fundamental behavior of metals and electronic phase transitions \citep{ElectronGas}. 
Several seminal \textit{ab initio} works have reported the energy of HEG at different densities \citep{ElectronGas, heg_method_3,heg_method_5, wapnet, ferminetGas}, and 
recently more investigations have been conducted using neural network ansatz \citep{wapnet,ferminetGas}. 
Here we broaden our tests to simulate a simple cubic cell containing 54 electrons in closed-shell configuration (Fig.~\ref{fig:unify}g).
Fig.~\ref{fig:unify}h shows the correlation energy error from our neural network calculations on HEG at 6 different densities from $r_s = 0.5 $ Bohr to 20.0 Bohr. 
The state-of-the-art results, namely VMC with backflow correlation (BF) \citep{heg_method_3} and distinguishable cluster with double excitations (DCD) \citep{heg_method_5} are also plotted for comparison, and the most accurate BF-DMC result is used as the reference energy of correlation error. 
Overall, our neural network performs very well, with an error of less than 1\% in a wide range of density (see Supplementary Table 11). 
Generally, the correlation error increases as the density of HEG decreases when the correlation effects become larger, which also appears in DCD calculations.

\subsection{Electron density}
\begin{figure}[t]
    \centering
    \includegraphics[width=0.70\columnwidth]{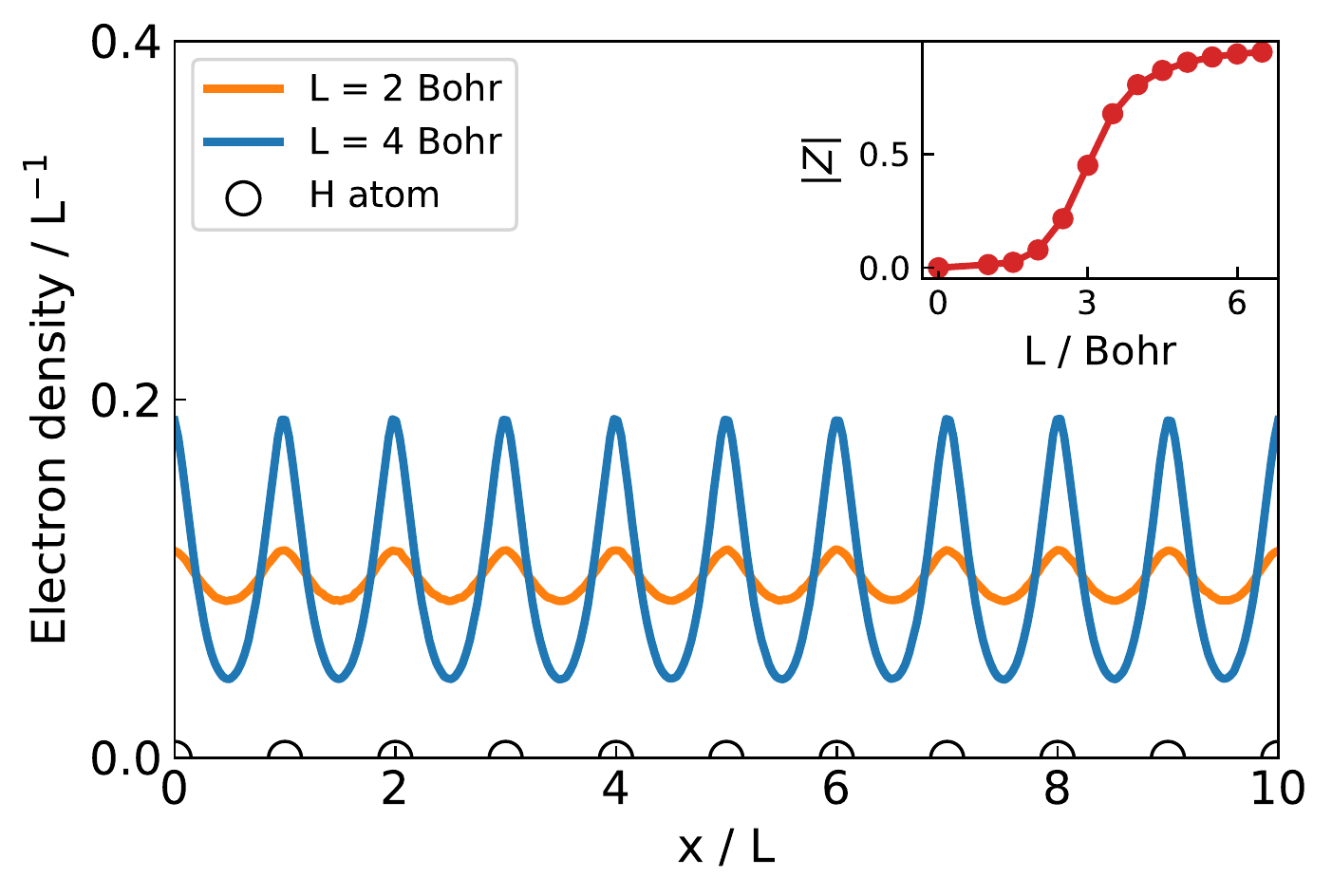}
    \caption{\textbf{Electron density of $\mathbf{{\rm H}_{10}}$ chains.}  Horizontal axis is scaled by the corresponding bond length. Complex polarization modulus ${|Z|}$ as a function of bond length is plotted in the inset. }
    \label{fig:h10_density}
\end{figure}

Besides the total energy of solid systems, the electron density is also a key quantity to be calculated.
For example, electron density is crucial for characterizing different solids, such as the difference between insulators and conductors, and the distinct nature of ionic and covalent crystals. 
In DFT the one-to-one correspondence between many-body wavefunction and electron density proved by Hohenberg and Kohn in 1964 suggests that electron density is a fundamental quantity of materials. 
However, 
an interesting survey found that while new functionals in DFT improve the energy calculation the obtained density somehow can deviate from the exact \citep{dft_science}. 
Here, with our accurate neural network wavefunction, we can also obtain accurate electron density of solids 
and provide a valuable benchmark and guidance for method development.


A conductor features free-moving electrons, which would have macroscopic movements under external electric fields. 
In contrast, electrons are localized and constrained in insulators and cause considerable electron resistance. 
In Fig.~\ref{fig:h10_density}, as an example, we show the calculated electron density of the hydrogen chain at two different bond lengths.
As we can see, for the compressed hydrogen chain (${\rm L} = 2$ Bohr), the electron density is rather uniform and has small fluctuations.
As the chain is stretched, the electrons are more localized and the density profile has larger variations.
The observation is consistent with the well-known insulator-conductor transition on the hydrogen chain by varying the bond length.
To measure the transition more quantitatively, we further calculate the complex polarization $Z$ as the order parameter for insulator-conductor transition \citep{vmc_complex_polarization}. 
A conducting state is characterized by a vanishing complex polarization modulus $|Z|\sim 0$, while an insulating state has finite $|Z|\sim1$. We can see that the conductor-insulator transition bond length of hydrogen chain is around 3 Bohr according to the calculated results, which is also consistent with previous studies \cite{vmc_complex_polarization}.

\begin{figure}[t]
    \centering
    \includegraphics[width=1.0\columnwidth]{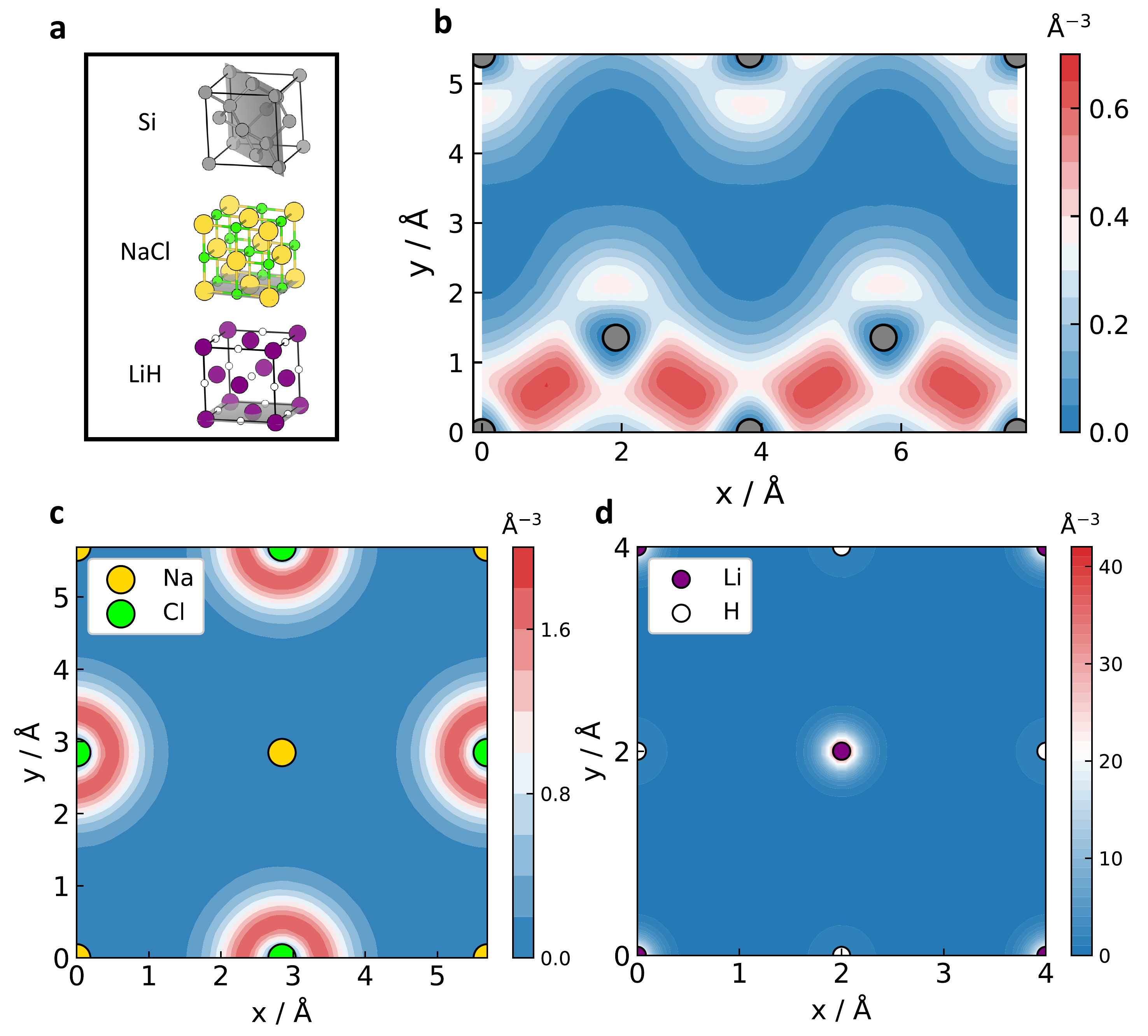}
    \caption{\textbf{Electron density of solids.} \textbf{a}, structures of solids, where the lattice planes for plotting electron densities are indicated. \textbf{b}, electron density of diamond-structured Si in its ($01\bar{1}$) plane, ccECP[Ne] is employed, and the bond length of Si equals $5.42 \AA$. \textbf{c}, electron density of NaCl crystal in its $xy$-plane, ccECP[Ne] is employed, and the bond length of NaCl equals $5.7 \AA$. \textbf{d}, electron density of LiH crystal in its $xy$-plane, and the bond length of LiH equals $4.0 \AA$.}
    \label{fig:density}
\end{figure}

Ionic and covalent bonds are the most fundamental chemical bonds in solids. While the physical pictures of these two types of bonding are very different, they both lie in the behavior of electrons as the "quantum glue" and electron density distribution is a simple way to visualize different bonding types. 
Here we choose to calculate the electron density of the diamond-structured Si, rock-salt NaCl and LiH crystals at their equilibrium position. They are representative of covalent and ionic crystals and have also been investigated by other high-level wavefunction methods, e.g. AFQMC \citep{afqmc_density}. 
Note that in the calculations of NaCl and Si, correlation-consistent effective core potential (ccECP) is employed to reduce the cost, which removes the inertia core electrons and keeps the behavior of active valence electrons \citep{ccecp, ferminetecp}. 

The electron density of diamond-structured Si in its $(01\bar{1})$ plane is plotted in Fig.~\ref{fig:density}b. We can see that valence electrons are shared by nearest Si atoms, forming apparent Si-Si covalent bonds. In contrast, valence electrons are located around atoms in NaCl crystal as Fig.~\ref{fig:density}c shows. All the valence electrons are attracted around Cl atoms, forming effective $\rm{Na^+}$ and $\rm{Cl^-}$ ions in the crystal. Moreover, the electron density of LiH crystal is also calculated and plotted in Fig.~\ref{fig:density}d. LiH crystal is a moderate system between a typical ionic and covalent crystal.
According to the result, the electrons are nearly equally distributed near Li and H atoms for our network. Detailed Bader charge analysis \citep{bader_charge} manifests the atoms in the crystal become ${\rm Li^{0.67+}}$ and ${\rm H^{0.67-}}$ ions respectively (resolution $\sim {\rm 0.015 \AA}$), which slightly deviates from the stable closed-shell configuration (see Supplementary Note 6 for more details).

\section{Conclusion}
The construction of a wave function for solid systems is a crucial but unsolved problem in the neural network community. 
The core mechanism of our neural network is the use of the periodic distance feature, which promotes molecule neural networks elegantly to the corresponding periodic ones and avoids time-consuming lattice summation. 
Considering the high-accuracy results obtained in this work, our neural network can be further applied to study more delicate physics and materials problems, such as the phase transitions of solids, surfaces, interfaces, and disordered systems, to name just a few.
Our ansatz also offers a flexible extension to other neural networks and an easy integration into traditional computational techniques.  
The naturally evolved many-body wavefunction from the neural network may provide more physical and chemical insights to emergent phenomena of complex materials. 

\section{Methods}  

\paragraph{Supercell approximation.}
Simulating a solid system requires solving the Schr\"{o}dinger equation of many electrons within a large bulk. Supercell approximation is usually adopted to simplify the problem, 
considering a finite number of electrons and nuclei with periodic boundary conditions, whose Hamiltonian reads
\begin{equation}
\begin{gathered}
    \hat{H}_S=\sum_i -\frac{1}{2}\Delta_i + \frac{1}{2}\sum_{\bL_S,i,j}'\frac{1}{|\br_i-\br_j+\bL_S|} \\
        - \sum_{\bL_S,i,I}\frac{Z_I}{|\br_i-\bR_I+\bL_S|} + \frac{1}{2}\sum_{\bL_S,I,J}'\frac{Z_I Z_J}{|\bR_I-\bR_J+\bL_S|}\ ,
\end{gathered}\label{eq:supercell_h}
\end{equation}
where $\br_i$ denotes the spatial position of i-th electron in the supercell. $\bR_I,Z_I$ are the spatial position and charge of I-th nucleus and $\{\bL_S\}$ is the set of supercell lattice vectors, which is usually a subset of primitive cell lattice vectors $\{\bL_p\}$. 
In order to simulate the real environments of electrons in solids, the interactions between the particles and their images are also included in $\hat{H}_S$, and the prime symbol in summation means $i=j$ terms are omitted for $\bL_S=0$.

Supercell Hamiltonian $\hat{H}_S$ is invariant under translation of any electron by a vector in $\{\bL_S\}$ as well as a simultaneous translation of all electrons by a vector in $\{\bL_p\}$. As a consequence, two periodic conditions are required for the ground-state wavefunction $\Psi$,
\begin{equation}
\label{eq:periodic_con}
\begin{gathered}
    \Psi(\br_1+\bL_p,...,\br_N+\bL_p)=\exp({i\bk_p\cdot\bL_p})\Psi(\br_1,...,\br_N)\ , \\
    \Psi(\br_1+\bL_S,...,\br_N)=\exp({i\bk_S\cdot\bL_S})\Psi(\br_1,...,\br_N)\ ,
\end{gathered}
\end{equation}
where $\bk_S, \bk_p$ denote the momentum vectors reduced in the first Brillouin zone of the supercell and the primitive cell, respectively. Eq.~\eqref{eq:periodic_con} and the anti-symmetry condition together form the fundamental requirements for $\Psi$. As the size of supercell increases, simulation results gradually converge to the thermodynamic limit of real solid system.

\paragraph{Wavefunction ansatz.} 
In conventional QMC simulation of solids, Hartree-Fock type wavefunction anzatz composed of Bloch functions is often used, which reads  
\begin{equation}
\begin{gathered}
\label{eq:hf}
    \Psi^{\rm HF}_{\bk_S,\bk_p}(\br)={\rm Det}\left|
    \begin{matrix}
    e^{i\bk_1\cdot\br_1}u_{\bk_1}(\br_1)  & \cdots & e^{i\bk_N\cdot\br_1}u_{\bk_N}(\br_1) \\
    \cdot & & \cdot\\
    \cdot & & \cdot\\
    \cdot & & \cdot\\
    e^{i\bk_1\cdot\br_N}u_{\bk_1}(\br_N)  & \cdots & e^{i\bk_N\cdot\br_N}u_{\bk_N}(\br_N) \\
    \end{matrix}
    \right|\ .
\end{gathered}
\end{equation}
In order to satisfy Eq.~\eqref{eq:periodic_con}, $\bk_i$ in the determinant should lie on the grid of supercell reciprocal lattice vectors $\{\bG_S\}$ offset by $\bk_S$ within the first Brillouin zone of primitive cell. Moreover, $u_{\bk}$ functions in Eq.~\eqref{eq:hf} should satisfy the translation invariant condition by the primitive cell lattice vectors,
\begin{equation}
\begin{gathered}
    \label{eq:uk_con}
    u_{\bk}(\br+\bL_p)=u_{\bk}(\br)\ .
\end{gathered}
\end{equation}
Following the strategy of FermiNet \citep{FermiNet}, Bloch functions in Eq.~\eqref{eq:hf} can be promoted with collective distances,
\begin{equation}
\begin{gathered}
\label{eq:promotion}
e^{i\bk\cdot\br_i}u_{\bk}(\br_i)\rightarrow e^{i\bk\cdot\br_i}u_\bk(\br_i;\br_{\neq i})\ ,
\end{gathered}
\end{equation}
where $\br_{\neq i}$ denotes all the electron coordinates except $\br_i$. 
These collective orbitals are constructed to achieve the equivalence of electron permutations $P$,
\begin{equation}
    P_{i,j}u_{\bk_i}(\br_j;\br_{\neq j})=u_{\bk_j}(\br_i;\br_{\neq i})\ ,
    \label{eq:permutation}
\end{equation}
which combined with the Slater determinant ensure the anti-symmetry nature of electron.
Moreover, we use the periodic distance features $d(\br)$ in Eq.~\eqref{eq:periodic_metric} to substitute ordinary $|\br|$ in the molecular neural network.
The periodic functions $f,g$ used in Eq.~\eqref{eq:periodic_metric} read
\begin{equation}
\begin{gathered}
\label{eq:fg}
    f(\omega)=|\omega|~(1-\frac{|\omega/\pi|^3}{4})\ ,\\
    g(\omega)=\omega~(1-\frac{3}{2}|\omega/\pi|+\frac{1}{2}|\omega/\pi|^2)\ ,
\end{gathered}
\end{equation}
and their arguments $\omega$ are reduced into $[-\pi,\pi]$.
Eq.~\eqref{eq:uk_con} can then be satisfied without causing discontinuity \citep{PeriodicDis}. For an overall sketch of the neural network, see Algorithm~\ref{alg:solidnet}. Note that the distance between electrons and nuclei is omitted for HEG system since it does not contain any nucleus. Specific hyperparameters of each system are listed in Supplementary Note 1.

\begin{algorithm}[h]
	\caption{\textbf{Pseudocode of network}} 
	\label{alg:solidnet}
	\begin{algorithmic}[1]
      \Require electron positions $\{\mathbf{r}_1^\uparrow, \cdots, \mathbf{r}_{n^\uparrow}^\uparrow, \mathbf{r}_1^\downarrow, \cdots, \mathbf{r}_{n^\downarrow}^\downarrow\}$
      \Require nuclear positions $\{\mathbf{R}_I\}$ in the primitive cell
      \Require lattice vector $\{\mathbf{a}^{p,S}_1,\mathbf{a}^{p,S}_2,\mathbf{a}^{p,S}_3\}$ of primitive cell and supercell
      \Require reciprocal lattice vector $\{\mathbf{b}^{p,S}_1,\mathbf{b}^{p,S}_2,\mathbf{b}^{p,S}_3\}$ of primitive cell and supercell
      \Require occupied $\{\mathbf{k_i}\}$ points offered by Hartree-Fock method
    \For{each electron e, atom I}
    \State $\omega_{e,I}=(\mathbf{r}_e-\mathbf{R}_I)\cdot\{\mathbf{b}^{p}_1,\mathbf{b}^{p}_2,\mathbf{b}^{p}_3\}$
    \State $\omega_{e,e'}=(\mathbf{r}_e-\mathbf{r}_{e'})\cdot\{\mathbf{b}^{S}_1,\mathbf{b}^{S}_2,\mathbf{b}^{S}_3\}$
    \EndFor
    \For {each electron e}
        \State $\mathbf{h}_e=\{\Sigma_{i=1}^3g(\omega_{e,I}^i)\ \mathbf{a}^p_i, d(\omega_{e,I})\}$
        \State $\mathbf{h}_{e,e'}=\{\Sigma_{i=1}^3g(\omega_{e,e'}^i)\ \mathbf{a}^S_i, d(\omega_{e,e'})\}$
        \EndFor
    \For{each layer l}
    \State $\mathbf{g}^{l,\uparrow}=\frac{1}{n^\uparrow}\sum_e\mathbf{h}_{e}^{l,\uparrow}$
    \State $\mathbf{g}^{l,\downarrow}=\frac{1}{n^\downarrow}\sum_e\mathbf{h}_{e}^{l,\downarrow}$
        \For{each electron e, spin $\alpha$}
            \State $\mathbf{g}^{l,\alpha,\uparrow}_e=\frac{1}{n^\uparrow}\sum_{e'}\mathbf{h}_{e,e'}^{l,\alpha,\uparrow}$
            \State $\mathbf{g}^{l,\alpha,\downarrow}_e=\frac{1}{n^\downarrow}\sum_{e'}\mathbf{h}_{e,e'}^{l,\alpha,\downarrow}$
            \State $\mathbf{f}^{l,\alpha}_e={\rm cat}(\mathbf{h}_e^{l,\alpha},\mathbf{g}^{l,\uparrow},
            \mathbf{g}^{l,\downarrow}, \mathbf{g}_e^{l,\alpha,\uparrow},
            \mathbf{g}_e^{l,\alpha,\downarrow})$
            \State $\mathbf{h}_e^{l+1,\alpha}={\rm tanh}(\mathbf{V}^l\cdot \mathbf{f}_e^{l,\alpha}+\mathbf{b}^l)+\mathbf{h}_e^{l,\alpha}$
            \State $\mathbf{h}_{e,e'}^{l+1,\alpha,\beta}={\rm tanh}(\mathbf{W}^l\cdot \mathbf{h}_{e,e'}^{l,\alpha,\beta}+\mathbf{c}^l)+\mathbf{h}_{e,e'}^{l,\alpha,\beta}$
        \EndFor
    \EndFor
    \For {each orbital i}
        \For{each electron e, spin $\alpha$}
            \State $u_{i,e}^\alpha={\rm Orb}_{i,\alpha}^{\rm Re}\cdot \mathbf{h}_e^L + \mathbf{i}\times{\rm Orb}_{i,\alpha}^{\rm Im}\cdot\mathbf{h}_e^L$
            \State $p_{i,e}^\alpha = \exp(\mathbf{i}\mathbf{k}_i\cdot\mathbf{r}_e^\alpha)$
            \State ${\rm enve}_{i,e}^\alpha=\sum_I \pi_{i}^{I,\alpha}\exp(-\sigma_{i}^{I,\alpha} d(\omega_{e,I}))$
            \State $\phi_{i,e}^\alpha=p_{i,e}^\alpha u_{i,e}^\alpha {\rm enve}^\alpha_{i,e}$
        \EndFor
    \EndFor
    \State $\Psi={\rm Det}[\phi^{\uparrow}]{\rm Det}[\phi^{\downarrow}]$
	\end{algorithmic}
\end{algorithm}

\paragraph{Neural network optimization.} 
Parameters $\theta$ within the neural network can be optimized to minimize the energy expectation value $\langle E_l \rangle$, and the gradient $\nabla_\theta\langle E_l\rangle$ reads
\begin{equation}
\label{eq:energy_grad}
\begin{gathered}
    \nabla_\theta\langle E_l\rangle = {\rm Re}[\langle E_l\nabla_\theta\ln\Psi^*\rangle - \langle E_l\rangle\langle\nabla_\theta\ln\Psi^*\rangle]\ , \\
    E_l = \Psi^{-1}\hat{H}_S\Psi\ ,
\end{gathered}
\end{equation}
where $E_l$ denotes the local energy of neural network ansatz $\Psi$. Besides energy minimization, stochastic reconfiguration optimization \citep{SR} has also been widely adopted and proved to be much more efficient, whose gradient reads
\begin{equation}
    \label{eq:sr}
    \begin{gathered}
        {\rm Grad} = F^{-1}\nabla_\theta\langle E_l\rangle\ ,\\
        F_{ij} = {\rm Re}\Big[\langle\frac{\partial\ln\Psi^*}{\partial \theta_i}\frac{\partial\ln\Psi}{\partial \theta_j}\rangle-\langle\frac{\partial\ln\Psi^*}{\partial\theta_i}\rangle\langle\frac{\partial\ln\Psi}{\partial\theta_j}\rangle\Big]\ .
    \end{gathered}
\end{equation}
In this work, we adopt a modified KFAC optimizer, which approximates $F$ as
\begin{equation}
    \label{eq:kfac}
    \begin{aligned}
    F& ={\rm Re}\Big[\langle\frac{\partial\ln\Psi^*}{\partial {\rm vec}(W_l)}\frac{\partial\ln\Psi^T}{\partial {\rm vec}(W_l)}\rangle-\langle\frac{\partial\ln\Psi^*}{\partial{\rm vec}(W_l)}\rangle\langle\frac{\partial\ln\Psi^T}{\partial{\rm vec}(W_l)}\rangle\Big] \\
    &= {\rm Re}\Big[\langle(a_l\otimes e_l^*)(a_l\otimes e_l)^T\rangle-\langle(a_l\otimes e_l^*)\rangle\langle(a_l\otimes e_l)\rangle^T\Big] \\
    &\approx {\rm Re}\Big[\langle a_la_l^T\rangle \otimes \langle e_l^*e_l^T\rangle\Big]\ ,
    \end{aligned}
\end{equation}
where $W_l$ denotes the weight parameters of layer $l$, and vec means vectorized form. $a_l,e_l$ denote the activation and sensitivity of layer $l$ respectively. Note that activation $a_l$ is always real-valued, which explains the absence of conjugation of $a_l$ in the second line. The first term in the bracket of Eq.~\eqref{eq:kfac} is approximated as the Kronecker product of the expectation values, and the second term is omitted for simplification.

\paragraph{Twist average boundary condition.}

TABC is a conventional technique to reduce the finite-size error due to the constrained size of supercell \citep{twist_average}. It averages the contributions from different periodic images of the supercell and improve the convergence on the total energy. The formula reads
\begin{equation}
\begin{gathered}
    E_{\rm TABC}=\frac{\Omega_S}{(2\pi)^3}\int_{\rm 1. B.Z.} d^3\bk_{S}~\frac{\Psi^*_{\bk_S}\hat{H}_S\Psi_{\bk_S}}{\Psi^*_{\bk_S}\Psi_{\bk_S}}\ ,
\end{gathered}
\end{equation}
where ${\rm 1. B. Z.}$ denotes the first Brillouin zone of supercell and the integral is practically approximated by a discrete sum of a Monkhorst-Pack mesh (see Supplementary Note 3.3 for more details). 

\paragraph{Structure factor correction} Finite-size error can be further reduced via the structure factor $S(\bk)$ correction \citep{sf_correction}, which is usually calculated to correct the exchange-correlation potential $V_{\rm xc}$ and the formula reads
\begin{equation}
\begin{gathered}
    \frac{\Delta V_{\rm xc}}{N_{\rm e}}=\frac{2\pi}{\Omega_S}\lim_{\bk\rightarrow 0}\frac{S(\bk)}{\bk^2}, \\ S(\bk)=\frac{1}{N_{\rm e}}[\langle\rho(\bk)\rho^*(\bk)\rangle-\langle\rho(\bk)\rangle \langle\rho^*(\bk)\rangle] \ ,\ 
\end{gathered}
\end{equation}
where $\lim_{\bk\rightarrow 0}$ is practically estimated via interpolation (see Supplementary Note 3.4 for more details).

\paragraph{Empirical correction formula.}

Empirical formulas are also commonly employed to reduce the finite-size error \citep{FoulkesReview}, one of which reads
\begin{equation}
\begin{gathered}
    E_\infty=E_{\rm N}^{\rm Net}+(E_{\infty}^{\rm HF}-E_{\rm N}^{\rm HF})\ .
    \label{eq:fneq}
\end{gathered}
\end{equation}
The simulation size of high-accuracy methods is usually limited due to high computational costs. Hence methods with much more practical time scaling, such as HF, is usually used to give a posterior estimation of the finite-size error. All the results of LiH are corrected using this empirical formula with HF results in ccpvdz basis (see Supplementary Note 4.3 for more details).  

\paragraph{Electron density analysis}
Electron density $\rho(\br)$ is defined as 
\begin{equation}
    \rho(\br)=N\int d^3\br_2\cdots d^3\br_N \ |\Psi(\br,\br_2,\cdots,\br_N)|^2,
    \label{eq:density}
\end{equation}
and it's practically evaluated by accumulating Monte Carlo samples of electrons on a uniform grid over the simulation cell. As for the complex polarization $Z$, it is defined as \citep{vmc_complex_polarization}
\begin{equation}
    Z = \langle \exp(i\sum_i\frac{2\pi}{L}\br_i^{\parallel})\rangle,
\end{equation}
where $\br^\parallel$ denotes the projection of electron coordinate along the chain direction. Moreover, Bader charge is employed to estimate the charge partition on each atom \citep{bader_charge}.
The convergence test of Bader charge is shown in the Supplementary Fig.8.

      


\paragraph{Workflow and computational details.}
This work is developed upon open-source FermiNet and PyQMC on Github, deep learning framework JAX \citep{jax2018github} is used which supports flexible and powerful complex number calculation. 
Ground-state energy calculations are performed with all-electrons. Diamond-structured Si and NaCl crystal are simulated with ccECP[Ne] \citep{ccecp}.
The neural network is pretrained by Hartree-Fock ansatz, obtained with
PySCF software \citep{pyscf}.
All the expectation values for distribution $|\Psi|^2$ are evaluated via the Monte Carlo approach, and then the energy and wavefunction is optimized using the modified KFAC optimizer (see Supplementary Fig.1, 2, 4, 6, 7). 
The Ewald summation technique is implemented for the lattice summation in energy calculation. 
After training is converged, energy is calculated in a separate inference phase. Concrete code of this work is developed on Github at \url{https://github.com/bytedance/DeepSolid}.




\begingroup
\setlength\bibsep{0pt}
\newcommand{\mathsl}{\mathit}
\footnotesize
\bibliography{ref}
\endgroup

\subsection*{Acknowledgments}

\begingroup
\footnotesize
The authors thank Matthew Foulkes, David Ceperley, Lucas Wagner, Gareth Conduit, and Ke Liao for helpful discussions. We thank ByteDance AML team specially for their technical and computing support. We also thank ByteDance AI-Lab LIT Group and the rest of ByteDance AI-Lab research team for inspiration and encouragement. This work is directed and supported by Hang Li and ByteDance AI-Lab. J.C. is supported by the National Natural Science Foundation of China under Grant No. 92165101.


\endgroup

\subsection*{Author contributions}

\begingroup
\footnotesize
X.L. and J.C. conceived the study; X.L. developed the method, performed implementations, simulations, and data analyses; Z.L. contributed to the code development and simulation of HEG; J.C. supervised the project. X.L., Z.L., and J.C. wrote the paper.
\endgroup

%

\end{document}


\title{Supplementary information: \textit{Ab initio} calculation of real solids via neural network ansatz}

\maketitle
\section{Hyperparameters for simulations}

The recommended hyperparameters are listed in Supplementary Table~\ref{tab:ge_hyper}. Some employed hyperparameters of the presented results differ from the recommended ones, which are specially given in Supplementary Table~\ref{tab:hyper}.
\begin{table}[htb]
\centering
\begin{tabular}{lclc}
\toprule
Hyperparameter & Value & Hyperparameter & Value  \\
\midrule
Pretrain basis & ccpvdz & Pretrain iterations  & 1e3\\  
Dimension of one electron layer $\mathbf{V}$ & 256 & Dimension of two electron layer $\mathbf{W}$ & 32 \\
Number of layers  & 4 & Number of determinants & 8\\
Optimizer & KFAC & Learning rate & 3e-2\\
Damping & 1e-3 & Constrained norm of gradient & 1e-3 \\
Momentum of optimizer & 0.0 & Batch size & \num{4096} \\
Number of training steps & 2e5 & Clipping window of gradient & 5 \\
MCMC burn in & 1e3 & MCMC steps between each iterations & 20 \\
MCMC move width & 2e-2 & Target MCMC acceptance & 55\% \\
Precision & Float64 &  Number of inference steps & 5e4 \\  
\bottomrule
\end{tabular}
\caption{\textbf{Recommended hyperparameters}
}\label{tab:ge_hyper}
\end{table}
\begin{table}[htb]
    \centering
    \begin{tabular}{|c|c|c|c|c|c|}\hline
    System & Layer dimension & Layer & Determinants & Batch size & Training steps \\\hline
    Hydrogen chain & (256, 32) & 3 & 8 & 4096 & 1e5\\
    Graphene & (256, 32) & 4 & 8 & 4096 & 3e5\\
    $2\times2\times2$ Lithium hydride&  (256, 32)& 4& 8 & 4096 & 3e5\\
    $3\times3\times3$ Lithium hydride & (256, 32)& 4& 1& 8192 & 4e5\\
    Homogeneous electron gas & (256, 32) & 3 & 1 & 4096 & 3e5 \\

    \hline
    \end{tabular}
        \caption{\textbf{Some system dependent hyperparameters}}
    \label{tab:hyper}
\end{table}
\section{Hydrogen chain}
\subsection{Training curve}
The training curve of ${\rm H}_{10}$ in PBC is plotted in Supplementary Fig.~\ref{fig:h10_training}. 
The correlation error is defined as 
\begin{equation}
    {\rm Correlation\ error} = \Big(1 - \frac{E_{\rm Net}-E_{\rm HF}}{E_{\rm DMC} - E_{\rm HF}}\Big) \times 100\%\ ,
\end{equation}
where $E_{\rm HF}$ is calculated using the ccpvdz basis set and $E_{\rm DMC}$ is taken from Ref.~\cite{hydrogen_chain}.

\begin{figure}[htb]
    \centering
    \includegraphics[width=0.9\columnwidth]{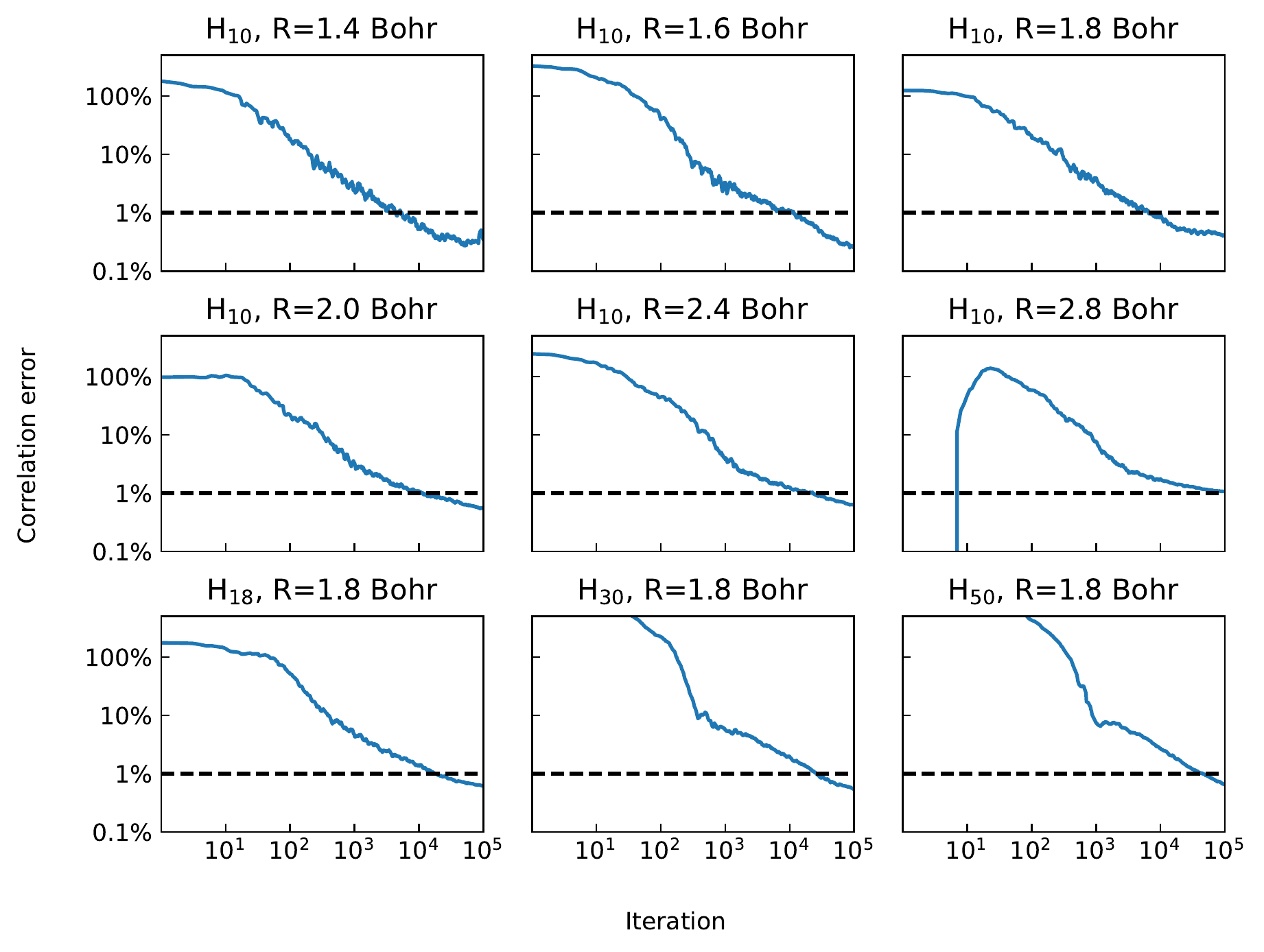}    \caption{\textbf{Hydrogen chain training curve.} For clarity, at each iteration number, we plot the median correlation error of the last 10$\%$ of the corresponding iteration.}
    \label{fig:h10_training}
\end{figure}

\subsection{${\rm H}_{10}$ dissociation curve}
Energy of ${\rm H}_{10}$ chain per atom is given in Supplementary Table~\ref{tab:h10}, LR-DMC and VMC results from Ref.~\citep{hydrogen_chain} are also listed for comparison.
\begin{table}[!h]
    \centering
    \begin{tabular}{|c|c|c|c|}\hline
    Bond length (${\rm \AA}$)& Net & LR-DMC(LDA) & VMC(LDA) \\\hline
    1.4 & -0.551677(1) & -0.55178(1) & -0.55049(1) \\
    1.6 & -0.568740(1) & -0.56881(1) & -0.56752(1) \\
    1.8 & -0.572922(1) & -0.57304(1) & -0.57172(1) \\
    2.0 &  -0.570401(1)& -0.57055(1)& -0.56911(1)\\
    2.4 & -0.556861(1) & -0.55703(1)& -0.55522(1)\\
    2.8 & -0.540783(1) & -0.54102(1)& -0.53831(1)\\

    \hline
    \end{tabular}
    \caption{\textbf{Energy of ${\rm H}_{10}$ chain}.}
    \label{tab:h10}
\end{table}

\subsection{Finite-size error extrapolation}
Energies of different hydrogen chains are given in Supplementary Table~\ref{tab:hn}.
\begin{table}[!h]
    \centering
    \begin{tabular}{|c|c|c|c|}\hline
    Size & Net & LR-DMC(LDA) & VMC(LDA) \\\hline
    10 & -0.572922(1) & -0.57304(1) & -0.57172(1) \\
    18 & -0.567776(1) & -0.56796(1) & -0.56644(1)\\
    30 & -0.566114(1) & -0.56627(1) & -0.56478(1)\\
    50 & -0.565419(1) & -0.56560(1) & -0.56409(1)\\
    \hline
    \end{tabular}
    \caption{Energies of different hydrogen chains, energies are given in Hartree and the bond length of hydrogen chain is fixed at 1.8 Bohr.}
    \label{tab:hn}
\end{table}

\section{Graphene}

\begin{table}[!h]
    \centering
    \begin{tabular}{|c|c|c|c|}\hline
    Atom & Position  (${\rm \AA}$) & Lattice vector & Position (${\rm \AA}$) \\\hline
    C1 & (1.421,   0.0,   0.0) & $\mathbf{a}_1$ &  (2.1315, -1.2306, 0.0)\\
    C2 & (2.842,   0.0,   0.0)  & $\mathbf{a}_2$ & (2.1315, 1.2306, 0.0)\\
     & & $\mathbf{a}_3$ & (0, 0, 52.9177)\\
    \hline
    \end{tabular}
    \caption{\textbf{Geometry of Graphene}}
    \label{tab:graphene}
\end{table}

\subsection{Geometry}
The primitive cell lattice vectors as well as carbon atom coordinates are given in Supplementary Table~\ref{tab:graphene}. The size of supercell is $2\times 2$.

\subsection{Twist average boundary condition (TABC)}
A $3\times3$ Monkhorst-Pack mesh in the first Brillouin zone of the supercell reciprocal space with $\Gamma$ point centered is used to approximate the twist average integral, which reads
\begin{equation}
\begin{gathered}
        E_{\rm TABC}=\frac{\Omega_S}{(2\pi)^3}\int_{\rm 1. B.Z.} d^3\mathbf{k}_{S}~\frac{\Psi^*_{\mathbf{k}_S}\hat{H}_S\Psi_{\mathbf{k}_S}}{\Psi^*_{\mathbf{k}_S}\Psi_{\mathbf{k}_S}} \approx \frac{1}{9} E_{\rm \mathbf{k}_1} + \frac{2}{3}E_{\rm \mathbf{k}_2} + \frac{2}{9}E_{\rm \mathbf{k}_3}, \\
        \mathbf{k}_1 = 0,\  
        \mathbf{k}_2 = \frac{1}{3} \mathbf{b}_1^S + \frac{1}{3} \mathbf{b}_2^S,\  
        \mathbf{k}_3 = \frac{2}{3} \mathbf{b}_1^S + \frac{1}{3} \mathbf{b}_2^S,
\end{gathered}
\end{equation}
and the weight factors origin from the different number of symmetry equivalent $\mathbf{k}$ points.

\subsection{Training curves}
Training curves at different $\mathbf{k}_S$ are plotted in Supplementary Fig.~\ref{fig:graphene_training}.

\begin{figure}[!h]
    \centering
    \includegraphics[width=0.9\columnwidth]{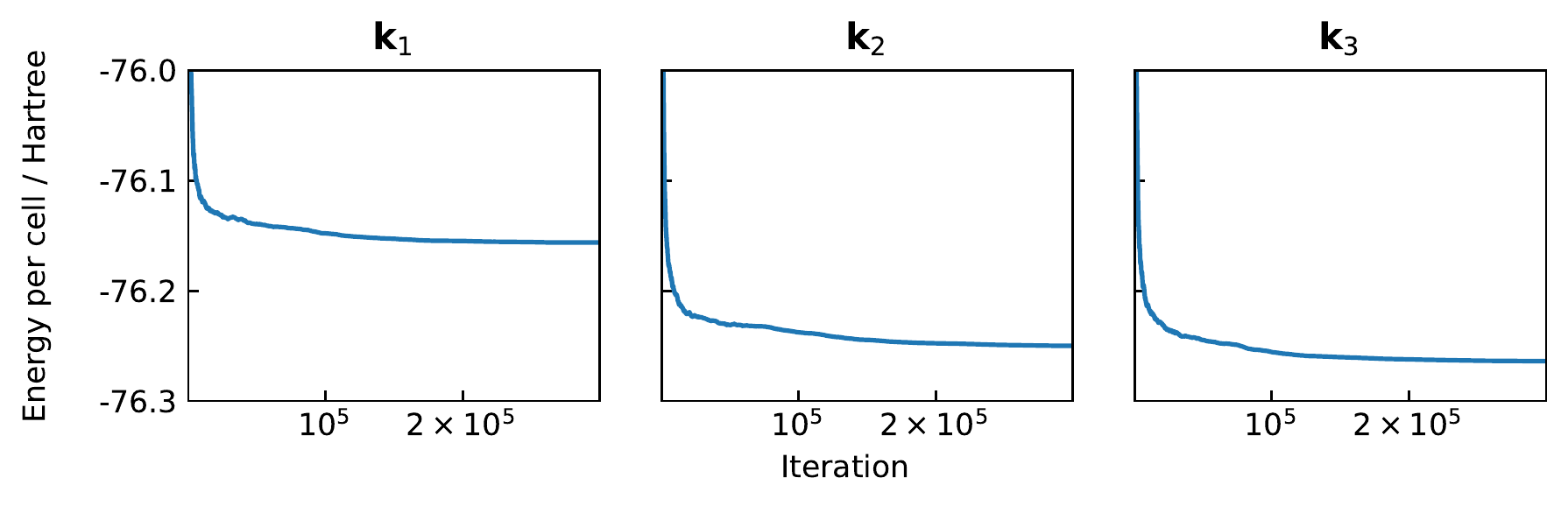}    \caption{\textbf{$2\times 2$ Graphene training curve.} For clarity, at each iteration number, we show the median energy per primitive cell over the last 10$\%$ of iteration.}
    \label{fig:graphene_training}
\end{figure}

The final results are listed in Supplementary Table~\ref{tab:graphene_energy}. 
The energy of an isolated carbon atom is taken from Ref.~\citep{FermiNet}, $E=-37.84471$ Hartree.
\begin{table}[!h]
    \centering
    \begin{tabular}{|c|c|c|c|}\hline
      & $\mathbf{k}_1$ & $\mathbf{k}_2$ & $\mathbf{k}_3$ \\\hline
    Energy (Hartree) & -76.15588(6) & -76.24949(5) & -76.26314(5) \\
    \hline
    \end{tabular}
    \caption{\textbf{Energy of graphene at different twists}}
    \label{tab:graphene_energy}
\end{table}

\subsection{Structure factor correction}
TABC technique is usually combined with structure factor corrections \citep{sf_correction}, and the combination is now seen as the standard scheme of applying QMC to solids. Structure factor $S(\bk)$ is calculated to correct the exchange-correlation part, namely $V_{\rm xc}$, of the total potential energy, which reads
\begin{equation}
\begin{gathered}
    \frac{\Delta V_{\rm xc}}{N_{\rm e}}=\frac{2\pi}{\Omega_S}\lim_{\bk\rightarrow 0}\frac{S(\bk)}{\bk^2}\ , \\ S(\bk)=\frac{1}{N_{\rm e}}[\langle\rho(\bk)\rho^*(\bk)\rangle-\langle\rho(\bk)\rangle \langle\rho^*(\bk)\rangle] \ ,\ 
    \rho(\bk)=\sum_i \exp(i\bk\cdot \br_i)\ ,
\end{gathered}
\end{equation}
where $\br_i$ refers to the coordinate of each electron, and $N_{\rm e}$ denotes the number of electrons in the simulation cell. The calculated $S(\bk)$ and corresponding $\Delta V_{\rm xc}$ of $\Gamma$ point is plotted in Supplementary Fig.~\ref{fig:graphene_sf}, and corrections of all twists are quite close to each other.

\begin{figure}[!h]
    \centering
    \includegraphics[width=0.9\columnwidth]{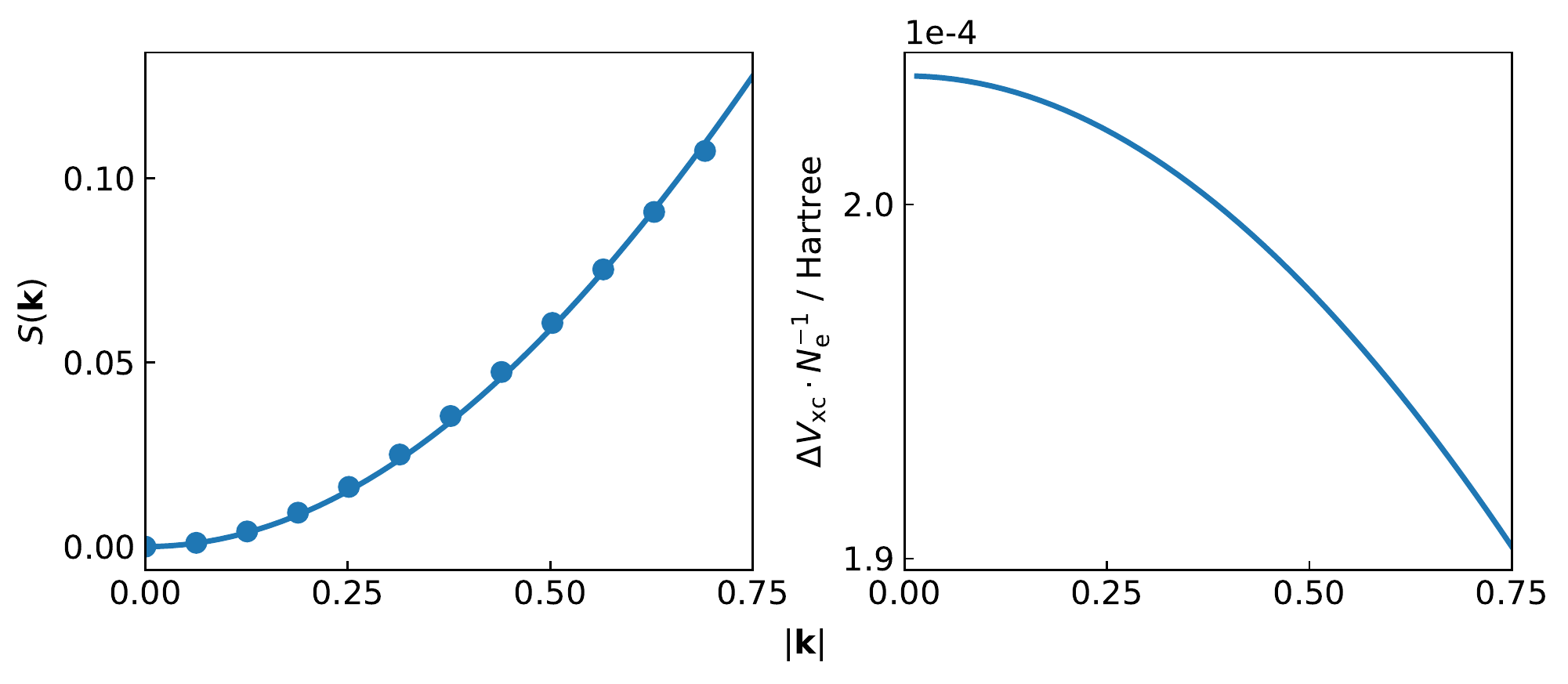}    \caption{\textbf{Structure factor correction of Graphene.} The lines are fitted with the formula: $S(\bk)=1-\exp(-a\cdot \bk^2)$.}
    \label{fig:graphene_sf}
\end{figure}
The final correction from structure factor is 0.00122 Hartree / atom.

\section{Lithium hydride}

\subsection{Geometry}
Lithium hydride crystal has a rock-salt structure, whose lattice vectors and atom positions are given in Supplementary Table~\ref{tab:lih}.

\begin{table}[!h]
    \centering
    \begin{tabular}{|c|c|c|c|}\hline
    Atom & Position   & lattice vector & Position  \\\hline
    Li & (0.0,   0.0,   0.0) & $\mathbf{a}_1$ &  (0.0, L/2, L/2)\\
    H & (L/2,   L/2,   L/2)  & $\mathbf{a}_2$ & (L/2, 0.0, L/2)\\
     & & $\mathbf{a}_3$ & (L/2, L/2, 0.0) \\
    \hline
    \end{tabular}
    \caption{\textbf{Geometry of LiH crystal}}
    \label{tab:lih}
\end{table}

\subsection{Training curves}
Training curves of the $2\times2\times2$ LiH crystal is plotted in Supplementary Fig.~\ref{fig:lih_training}.
\begin{figure}[!h]
    \centering
    \includegraphics[width=0.9\columnwidth]{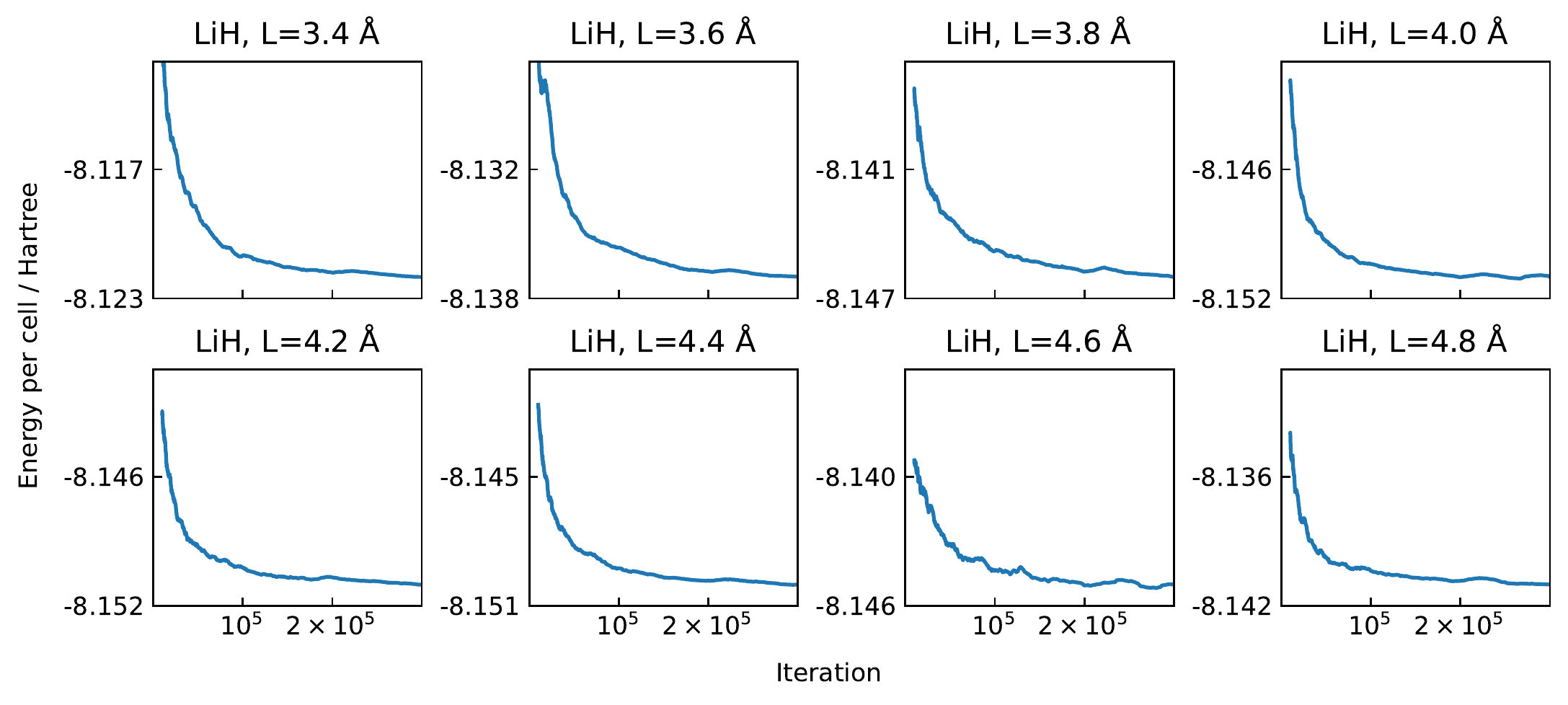}    \caption{\textbf{$2\times 2\times2$ LiH training curve.} For clarity, at each iteration number, we show the median energy of primitive cell over the last 10$\%$ of iteration.}
    \label{fig:lih_training}
\end{figure}

\subsection{Dissociation curve}
The energy of $2\times2\times2$ LiH is listed in Supplementary Table~\ref{tab:lih_energy}. 
The energy of an isolated lithium atom is taken from Ref.~\citep{FermiNet}, $E=-7.47798 ~{\rm Hartree}$. 
Corresponding Hatree-Fock corrections are calculated with the ccpvdz basis set and the convergence behavior of HF calculation is plotted in Supplementary Fig.~\ref{fig:hf_fn}.

\begin{table}[!h]
    \centering
    \begin{tabular}{|c|c|c|c|c|c|}\hline
    L (${\rm \AA}$)& Net & HF correction & L (${\rm \AA}$) & Net & HF correction  \\\hline
    3.4 & -8.12185(1) & -0.0099 & 4.2 & -8.15112(1) & -0.0004\\
    3.6 & -8.13738(1) & -0.0067 & 4.4 & -8.14967(1) & 0.0009\\
    3.8 & -8.146147(1) & -0.0042 & 4.6 & -8.14502(1) & 0.0020\\
    4.0 &  -8.15096(1)& -0.0021& 4.8 & -8.14094(1)& 0.0030\\
    \hline
    \end{tabular}
    \caption{\textbf{Energy of $2\times 2 \times 2$ LiH crystal.} Energies are all given in Hartree.}
    \label{tab:lih_energy}
\end{table}

\begin{figure}[!h]
    \centering
    \includegraphics[width=0.9\columnwidth]{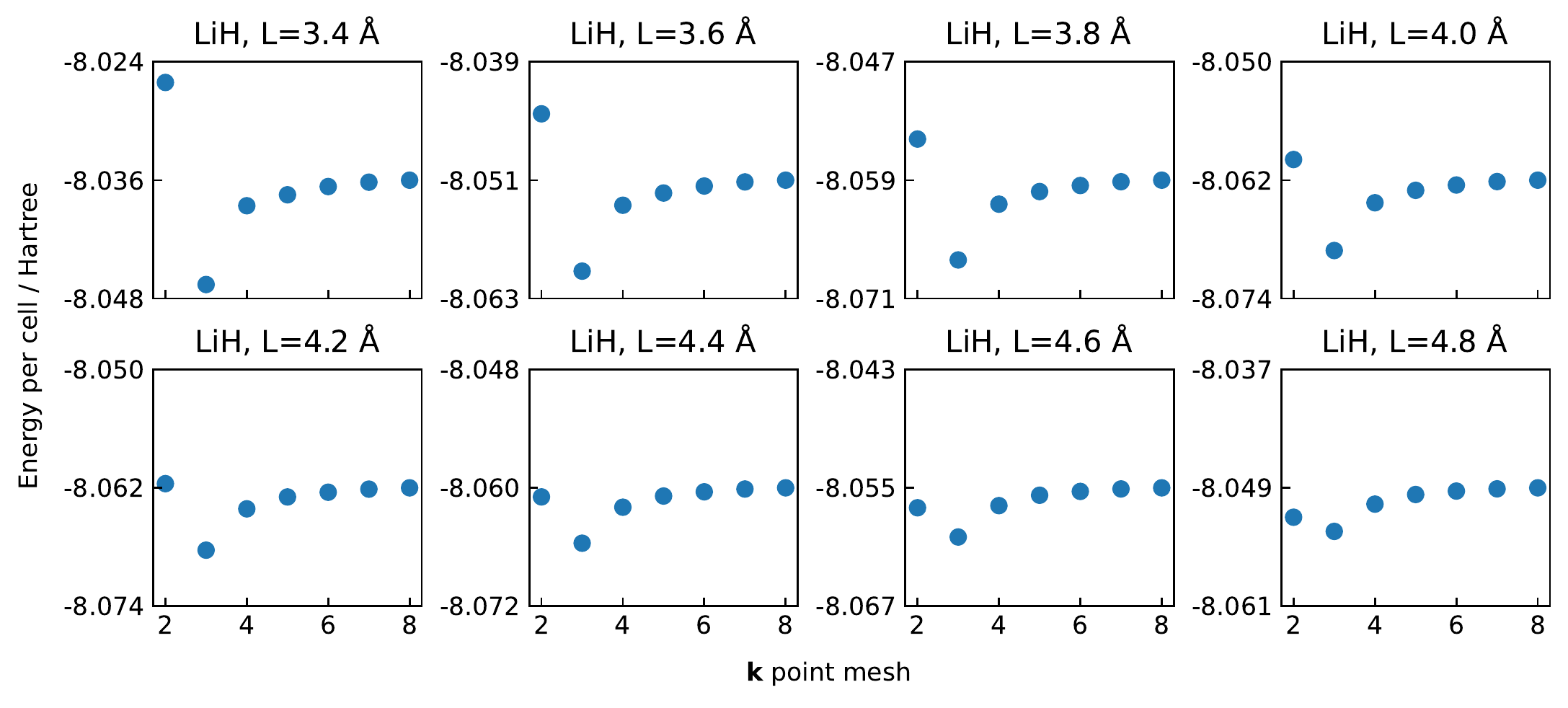}    \caption{\textbf{Hartree-Fock corrections.} The convergence behavior of HF calculations with respect to the number of $\mathbf{k}$ points.}
    \label{fig:hf_fn}
\end{figure}

\subsection{Birch-Murnaghan fit}
The third order Birch-Murnaghan equation of state is employed to fit the dissociation curve, which reads
\begin{equation}
    E(V) = E_0 + \frac{9V_0B_0}{16}\Big\{\Big[\Big(\frac{V_0}{V}\Big)^{2/3}-1\Big]^3B_0'+\Big[\Big(\frac{V_0}{V}\Big)^{2/3}-1\Big]^2\Big[6-4\Big(\frac{V_0}{V}\Big)^{2/3}\Big]\Big\},
\end{equation}
where $E_0,V_0,B_0,B_0'$ are fitted quantities, their results and corresponding experiment data \citep{qmc_lih} are listed in Supplementary Table~\ref{tab:bh_fit}.

\begin{table}[!h]
    \centering
    \begin{tabular}{|c|c|c|c|c|}\hline
     &$a_0$ (\AA) & $B_0$ (GPa)  & $E_{\rm coh} $ (eV) \\\hline
    Net & 4.022 & 36.89 & -4.757 \\
    Exp & 4.061(1) & 33-38 & -4.778,-4.759 \\
    \hline
    \end{tabular}
    \caption{\textbf{Parameters of Birch-Murnaghan equation of state}}
    \label{tab:bh_fit}
\end{table}

\subsection{$3\times3\times3$ LiH}
The training curve of the $3\times3\times3$ LiH crystal at its equilibrium lattice constant ${\rm L}=4.061{\rm \AA}$ is plotted in Supplementary Fig.~\ref{fig:lih_333_training}, corresponding Hartree-Fock corrections are also given. The final inference results from neural network are listed in Supplementary Table~\ref{tab:lih_333_energy}.
\begin{figure}[!h]
    \centering
    \includegraphics[width=0.9\columnwidth]{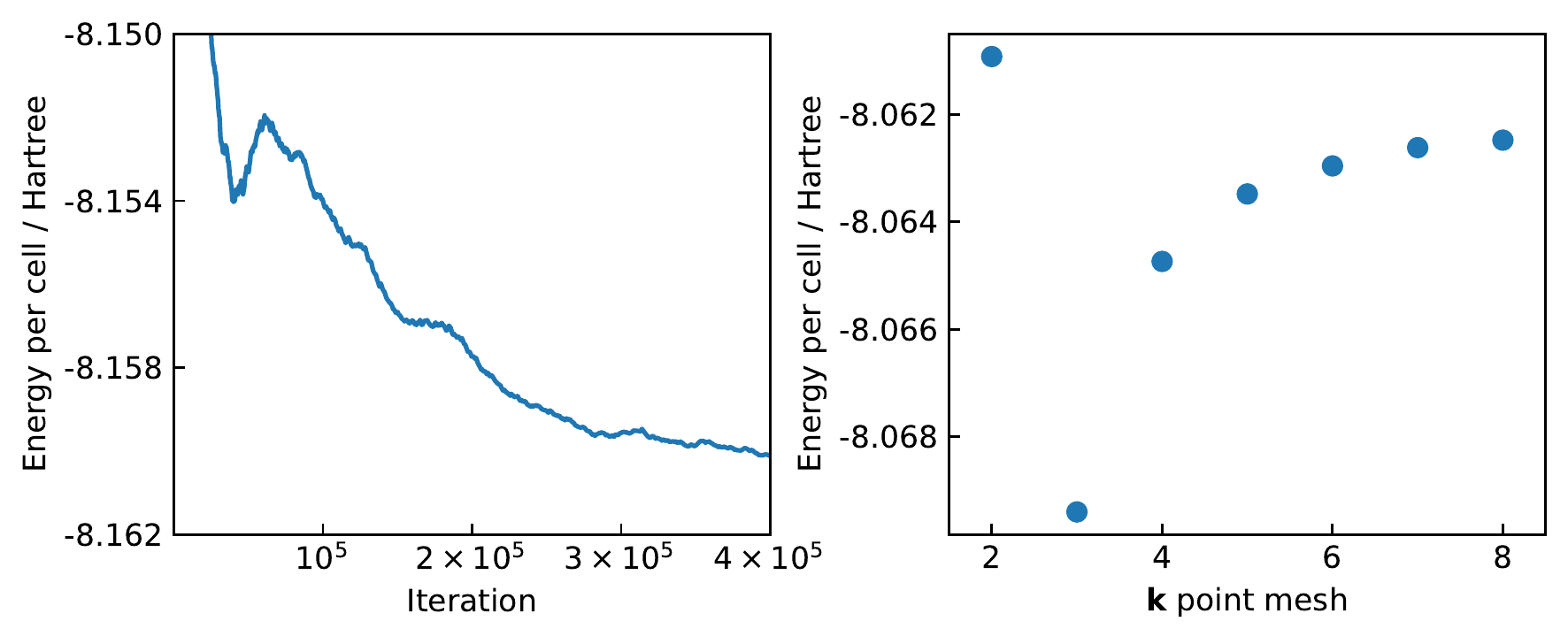}    \caption{\textbf{$3\times 3\times3$ LiH.} Left panel plots the training curve of the $3\times3\times3$ LiH. For clarity, at each iteration number, we show the median energy per unit cell over the last 10$\%$ of iteration. Right panel plots the corresponding Hartree-Fock corrections with the ccpvdz basis set.}
    \label{fig:lih_333_training}
\end{figure}
\begin{table}[!h]
    \centering
    \begin{tabular}{|c|c|c|}\hline
    L (${\rm \AA}$)& Net & HF correction  \\\hline
    4.061 & -8.16020(2) & 0.0069\\
    \hline
    \end{tabular}
    \caption{\textbf{Energy of the $3\times 3 \times 3$ LiH crystal.} Energies are all given in Hartree.}
    \label{tab:lih_333_energy}
\end{table}

\section{Homogeneous electron gas}

\subsection{Training curve}

The training curve of HEG system containing 54 electrons is plotted in Supplementary Fig.~\ref{fig:electron_gas_training}. 
$E_{\rm HF}$ and $E_{\rm DMC}$ are taken from Ref. \cite{heg_method_3}. 
Final results of neural network, BF-DMC, BF-VMC and DCD \citep{heg_method_3,heg_method_5} are listed in Supplementary Table~\ref{tab:electron_gas}.

\begin{figure}[!h]
    \centering
    \includegraphics[width=0.9\columnwidth]{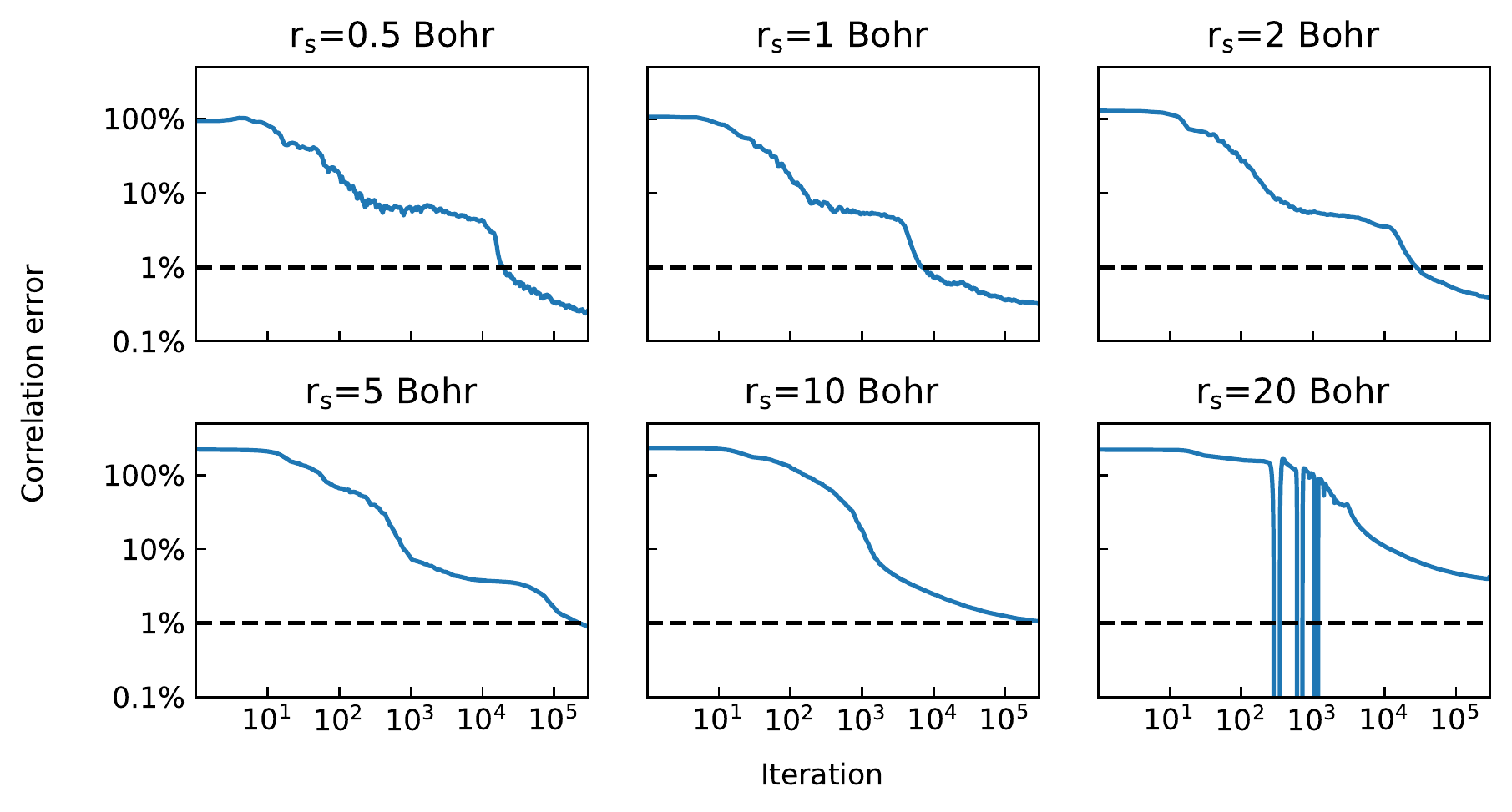}    \caption{\textbf{Homogeneous electron gas.} For clarity, at each iteration number, we show the median correlation error over the last 10$\%$ of iteration.}
    \label{fig:electron_gas_training}
\end{figure}

\begin{table}[!h]
    \centering
    \begin{tabular}{|c|c|c|c|c|}\hline
    $r_s$ & Net & BF-DMC & BF-VMC & DCD \\\hline
    0.5 & 3.221226(2) & 3.22112(4) & 3.22132(7) & 3.22052\\
    1 & 0.530019(1) & 0.52989(4) & 0.53009(3) & 0.53001\\
    2 & -0.013840(1) & -0.013966(9) & -0.01382(2) & -0.01286 \\
    5 & -0.0788354(2) & -0.079036(3) & -0.078961(5) & -0.07655\\
    10 & -0.0542785(1) & -0.054443(2) & -0.054389(2)& -0.05157\\
    20 & -0.0316886(1) & -0.032047(2) & -0.0319984(8)& -0.02925\\
    \hline
    \end{tabular}
    \caption{\textbf{Energy per electron of HEG at different mean radius of electrons $r_s$.} HEG system contains 54 electrons and energies are all given in Hartree, $r_s$ is given in Bohr.}
    \label{tab:electron_gas}
\end{table}

\section{Bader charge analysis}
\begin{figure}[!h]
    \centering
    \includegraphics[width=0.9\columnwidth]{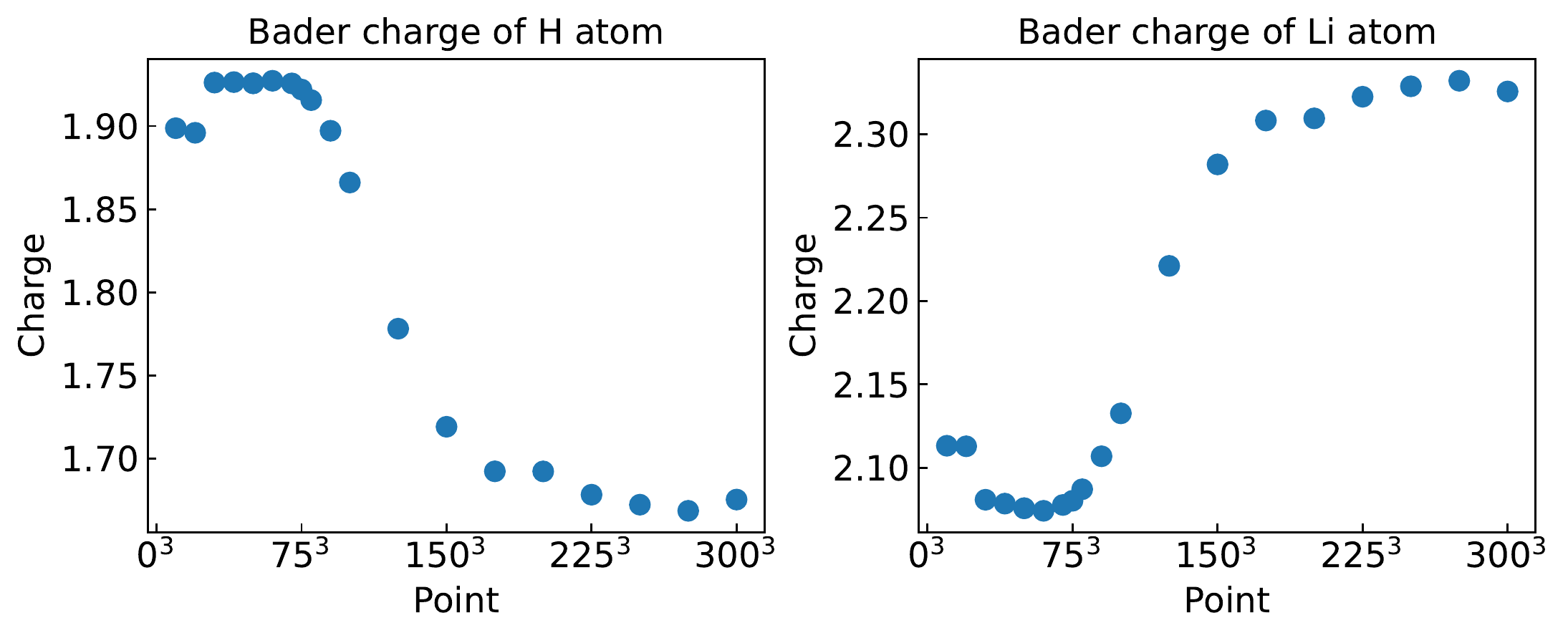}    \caption{\textbf{Bader charge of LiH crystal.} Calculated Bader charge of atoms in LiH crystal, point denotes the number uniformly dividing the crystal.}
    \label{fig:bader}
\end{figure}

Detailed Bader charge analysis \cite{bader_charge} is applied to conventional LiH crystal, and the result is plotted in Fig.~\ref{fig:bader}.

According to the result, Li and H atoms in LiH become ${\rm Li}^{0.67+}$ and ${\rm H}^{0.67-}$ ions respectively. 

\bibliography{supplement}